\documentclass[twocolumn]{revtex4}
\usepackage{graphicx,multirow,array,amsmath,empheq,amsthm,amssymb,amsfonts,tikz,ragged2e,color}
\begin{document}
\title{Propagation of Compressional Alfv\'en Waves in a Magnetized Pair Plasma Medium}
\author{T. I. Rajib* and S. Sultana}
\address{Department of Physics, Jahangirnagar University, Savar, Dhaka-1342, Bangladesh\\
*Corresponding Author's E-mail Address: tirajibphys@juniv.edu OR tirajibphys@gmail.com}
\begin{abstract}
The reductive perturbation approach was used to explore the nonlinear propagation of fast (compressive) and slow (rarefactive) electron-position (EP) magnetoacoustic (EPMA) modes in an EP plasma medium. The solitary wave solution of the Korteweg-de Vries (K-dV) equation is used to identify the basic properties of EP compressional Alfv\'{e}n waves. It is shown that the fast (slow) EPMA mode is predicted to propagate as compressive (rarefactive) solitary waves. The basic features (i.e., speed, amplitude, and width) of the compressive (i.e., fast) EPMA waves are found to be completely different from those of rarefactive (i.e., slow) EPMA ones. It is also examined that hump (dip) shape solitary waves are found for fast (slow) mode. The significance of our findings in understanding the nonlinear electromagnetic wave phenomena in laboratory plasma and space environments where EP plasma may exist.
\end{abstract}

\maketitle

\section{Introduction}
The electron-positron (EP) plasmas are observed in
various astrophysical environments like the solar wind \cite{Clem2000,Adriani2009,Adriani2011}, the magnetosphere of Earth \cite{Ackermann2012,Aguilar2014}, pulsars magnetosphere \cite{Profumo2012}, and microquasars \cite{Siegert2016}. In addition, this pair of plasmas can be
produced by imposing an ultra-intense laser on the matter in
the laboratory \cite{Sarri2013,Sarri2015}. It is also mentioned that EP plasmas are created in the presence
of strong magnetic/electric fields or extremely high
temperatures \cite{Thoma2009}. The progress of high-efficiency
techniques for assembling pure positron plasmas in Penning traps
\cite{Greaves1994,Surko1989} now makes laboratory experiments
able to produce this type of pair plasmas. It is shown that EP pair
production is expected to occur in post-disruption plasmas in
large tokamaks \cite{Helander2003}. The physics of an EP
plasma is quite different from electron-ion plasmas due to the
large ion-to-electron mass \cite{Swanson2003,Stix1992}. The EP
plasmas at the surface of magnetars and fast-rotating neutron stars are held in strong magnetic fields, while superseding
magnetic fields can be created in intense laser-plasma
interaction experiments. Therefore, the understanding of
collective phenomena in magnetized pair plasmas has been a topic
of significant interest \cite{Zank1995,Iwamoto1993,Lominadze1982,Gedalin1985,Yu1986,Brodin2007,Rajib2015}.

A vast amount of theoretical and numerical research has been
done, describing the unique physics of these plasmas,
including basic wave physics \cite{Zank1995}, recollection
\cite{Bessho2005,Blackman1994,Yin2008}, and nonlinear solitary waves \cite{Berezhiani1994,Cattaert2005}. Sakai and
Kawata \cite{Sakai1980} has analyzed the small amplitude solitary
waves by higher-order modified Korteweg-de Vries (mK-dV) equation from
relativistic hydrodynamic equations and examined the
nonlinear evolution of circularly polarized Alfv\'{e}n wave. On the other hand, Zank
and Greaves \cite{Zank1995} have presented the linear properties
of different electrostatic and electromagnetic modes in  both
unmagnetized and magnetized pair plasmas and investigated the properties of non-enveloped solitary waves.
El-Wakil \textit{et al.} \cite{Elwakil2019} studied the propagation of solitary waves and double-layers in EP pair plasmas with stationary ions and non-extensive electrons based on reductive perturbation theory via different nonlinear K-dV, mK-dV and Gardner’s equations.
Iwamoto \cite{Iwamoto1993} has discussed a kinetic description of
numerous linear collective modes in a non-relativistic pair
magnetoplasma. Helander and Ward \cite{Helander2003} have shown
that positrons can be created within tokamaks due to collisions
of runaway electrons with plasma ions/thermal electrons.
Lontano \textit{et al.} \cite{Lontano2001} have investigated the
interaction between arbitrary amplitude electromagnetic (EM)
fields and EP (hot) plasma. The nonlinear wave propagation of EP plasma in a pulsar magnetosphere has been investigated by
various authors \cite{Kennel1976,Arons1986} by applying various
theoretical approaches, they observed that large-amplitude
supercilious wave determines the average plasma properties. Many
theoretical investigations
\cite{Sakai1980,Melikidze1981,Stenflo1985} have been carried out
on the nonlinear Alfv\'{e}n waves propagation in an EP plasma.
Melikidze \textit{et al.} \cite{Melikidze1981} analyzed the
solitary waves properties for field-aligned electromagnetic waves
(via the cubic Schr\"{o}dinger equation) by employing a kinetic
approach. The authors \cite{Stenflo1985} considered Alfv\'{e}n
waves in the cold e-p plasma and described the coupling of the
radiation with cold electrostatic oscillation, while the authors
\cite{Mikhailovskii1985} have examined  the nonlinear
Alfv\'{e}n waves theoretically in a relativistic e-p plasma.

Apart from the opposite polarity EP pair plasma, recently, based on the theoretical predictions and
satellite/experimental observations, several authors
\cite{D'Angelo2001,D'Angelo2002,Shukla2006,Mamun2002,Sayed2007,Rahman2008,Mamun2008a,Mamun2008b,Verheest2009}
have considered a dusty plasma with the dust of opposite polarity, and
have investigated linear
\cite{D'Angelo2001,D'Angelo2002,Shukla2006} and nonlinear
\cite{Mamun2002,Sayed2007,Rahman2008,Mamun2008a,Mamun2008b,Verheest2009}
electrostatic waves excluding
\cite{D'Angelo2001,Shukla2006,Mamun2002,Sayed2007,Rahman2008,Mamun2008a,Mamun2008b,Verheest2009}
or including \cite{D'Angelo2002} external static magnetic field. On
the other hand, Shukla \cite{Shukla2004} and Mamun
\cite{Mamun2011} have considered a medium of magnetized opposite
polarity dust to study linear and nonlinear
electromagnetic waves. Shukla has analytically studied linear
dispersive dust Alfv\'en waves, and associated dipolar vortex
\cite{Shukla2004}, whereas Mamun has examined the nonlinear
propagation of the fast and slow dust magnetoacoustic (MA) modes
\cite{Mamun2011}.

In the earlier work \cite{Mamun2011}, Mamun studied the compressive and rarefactive electromagnetic solitary structures in an opposite
polarity dusty plasma medium. Mamun concluded that this plasma model is applied to any opposite polarity plasma medium such as EP and Electron-ion. This
research article deals with the extension of the earlier one \cite{Mamun2011} with the ratio of positron mass to electron mass is equal to unity i.e., $\alpha=m_p/m_e=1$. We, in our present work considers a medium of EP plasma, and made a systematic analysis of high-frequency nonlinear Alfv\'en solitary waves by using a well-known K-dV equation with reductive perturbation technique.

The layout of the manuscript is as follows: The governing equations describing our plasma model and a brief description of the mathematical technique, followed by the derivation of a K-dV equation with its solution
is given in Section \ref{GE}. The numerical analysis and results are presented in Section \ref{NA}, and finally, a discussion is drawn in Section \ref{D}.

\section{Governing Equations}
\label{GE}
We consider the propagation of electromagnetic perturbations in a
medium of EP, which is assumed to be immersed
in an external static magnetic field $\vec{B}_0$ lying in the
$y$-$z$ plane. Thus, the macroscopic state of the medium of
opposite polarity magnetized EP fluids
\cite{D'Angelo2002,Shukla2004,Mamun2011} is described by the following a set of equations
\begin{eqnarray}
&&\hspace*{5mm}\frac{\partial N_s}{\partial t} + {\vec\nabla}\cdot (N_s{\vec U}_s) = 0,\label{1}\\
&&\hspace*{5mm}m_sD_t^{s}{\vec U}_s =q_{s}\left({\vec E} +\frac{1}{c}{\vec U}_s\times{\vec B}\right)-{\vec\nabla}P_s,\label{2}\\
&&\hspace*{5mm}{\vec \nabla }\times {\vec E} = - \frac{1} {c}\frac{\partial {\vec B}} {\partial t},\label{3}\\
&&\hspace*{5mm}{\vec \nabla} \times{\vec B} = \frac{4 \pi}{c}\sum_{s} q_{s}  N_s{\vec U}_s+ \frac{1}{c}\frac{\partial{\vec E}}{\partial t},\label{4}\\
&&\hspace*{5mm}{\vec \nabla }\cdot{\vec E} = 4 \pi \sum_{s} q_{s} N_{s},\label{5}\\
&&\hspace*{5mm}{\vec \nabla}\cdot{\vec B} = 0, \label{6}
\end{eqnarray}
where $D_t^{s}=\partial_t +{\vec U}_s \cdot \vec \nabla$;
$s~(= e, p)$ denotes the species (namely, electron and
positron); $m_s$, $q_s$, and $N_s$ are, respectively, mass,
charge, and number density of the species $s$;  ${\vec U}_s$ is
the hydrodynamic velocity; $P_s = N_sT_s$ with $T_s$, the thermal energy; ${\vec E}$ is the electric field, and
${\vec B}$ is the magnetic field; $c~(=3\times10^8~m/s)$ is the speed of light in a
vacuum.
Now, neglecting the contribution of the displacement
current as of the wave phase speed is negligible compared to the
speed of light $c$ and assuming the quasi-neutrality condition
($N_e=N_p$), we can reduce our basic equations
(\ref{1})$-$(\ref{6}) to
\begin{eqnarray}
&&\hspace*{-1mm}{\partial_t n_s} + {\vec \nabla}\cdot(n_s{\vec u}_s) = 0,\label{7}\\
&&\hspace*{-1mm}n_sD_t^{s}\left[(1+\alpha){\vec u}_s +\frac{\vec{\alpha\cal H}}{n_s}\right]-{\vec{\cal H}\times{\vec b}}+\beta{\vec\nabla}n_s=0, \label{8}\\
&&\hspace*{-1mm}{\partial_t \vec b}-\vec\nabla\times\left[{\vec u}_s\times{\vec b}+D_t^{s}{\vec u}_s+\frac{\beta{\vec\nabla}n_s}{n_s}\right]=0,\label{9}
\end{eqnarray}
where  $n_s=N_s/N_{s0}$, $u_s=U_s/V_{As}$ with $V_{As}=B_0/\sqrt{4\pi N_{s0}m_s}$, $\vec b= \vec B/B_0$, $\vec {\cal H}=\vec \nabla\times\vec b$, $\alpha=m_p/m_n=1$ (Note that we consider $\alpha$ is equal to unity on the rest of the manuscript for our considered system as the mass of electron and positron is same), $\beta=P_T/P_B$ is the ratio of the net EP thermal pressure $P_T(=P_e+P_p)$ to the magnetic pressure $P_B=B_0^2/4\pi$. We note the from equation (\ref{7}), and thereafter, time variable is normalized by $\omega_{cs}^{-1}=m_sc/q_sB_0$, and space variable is normal by $\delta_s=V_{As}/\omega_{cs}$.

We are interested in high-frequency small but finite amplitude electromagnetic
perturbation modes propagating along the $z$-axis (i.e., all
dependent variables depend on $z$ and $t$ only) in the presence of an
external static magnetic field ${\vec B}_0$ that makes an angle
$\theta$ with the $z$-axis. We construct a weakly nonlinear theory
which leads to scaling of the independent variables
through the K-dV stretched coordinates \cite{Washimi1966,Haider2019}
\begin{eqnarray} \left.
\begin{array}{l}
\xi=\epsilon^{1/2}(z-V_0t),\\
\tau=\epsilon^{3/2}t,
\end{array}
\right\} \label{10}
\end{eqnarray}
where $V_0(=\omega/kV_{As})$ is the phase speed normalized by $V_{As}$, where $\omega$ is the wave frequency, $k$ is the propagation constant), and
 $\epsilon$ $(0<\epsilon<1)$ is a small parameter measuring the weakness of the dispersion. We then expand the dependent variables $n_s$, $u_{sx}$, $u_{sy}$, $u_{sz}$, $b_x$, and $b_y$ (where the subscript $x$, $y$, and $z$, respectively, represent
$x$-, $y$- and $z$-components of the quantity involved) about their equilibrium values in powers of
$\epsilon$:
\begin{eqnarray} \left.
\begin{array}{l}
n_s=1+\epsilon n_s^{(1)}+\epsilon^{2}n_s^{(2)}+ \cdot \cdot \cdot,\\
u_{sx}=0+\epsilon^{3/2} u^{(1)}_{sx}+\epsilon^{5/2}u_{sx}^{(2)}+\cdot \cdot \cdot,\\
u_{sy}=0+\epsilon u_{sy}^{(1)}+\epsilon^{2}u_{sy}^{(2)}+\cdot \cdot \cdot,\\
u_{sz}=0+\epsilon u^{(1)}_{sz}+\epsilon^{2}u_{sz}^{(2)}+ \cdot \cdot \cdot,\\
b_x=0+\epsilon^{3/2}b_x^{(1)}+\epsilon^{5/2}b_x^{(2)}+\cdot \cdot \cdot,\\
b_y=sin\theta+\epsilon b_y^{(1)}+\epsilon^{2}b_y^{(2)}+\cdot \cdot
\cdot,\\
b_z=cos\theta.
\end{array}
\right\} \label{11}
\end{eqnarray}

Now, substituting equations (\ref{10}) and (\ref{11}) into
equations (\ref{7})-(\ref{9}), and equating the coefficients of
$\epsilon^{3/2}$ we obtain a set of equations
that can be simplified as
\begin{eqnarray}
&&\hspace*{-1mm} u^{(1)}_{sz}=[{V_0sin\theta}/({2V_{0}^2-\beta})]B_{y}^{(1)},\label{12}\\
&&\hspace*{-1mm} n_s^{(1)}=[{sin\theta}/({2V_{0}^2-\beta})]B_{y}^{(1)},\label{13}\\
&&\hspace*{-1mm} u^{(1)}_{sy}=-[{cos\theta}/{2V_{0}}]B_{y}^{(1)},\label{14}\\
&&\hspace*{-1mm}V_{p}=\sqrt{\frac{1}{2}\bigg[1+\beta\pm\sqrt{(1+\beta)^2-4\beta{cos^2\theta}}\bigg]}, \label{15}\
\end{eqnarray}
where $V_p = \sqrt{2}V_0$ represents the pair MA wave phase speed
normalized by the effective pair-Alfv\'{e}n speed $V_A = P_B /\rho_T =
V_{As}/\sqrt{2}$ $(i.e., V_p = \omega/kV_A)$, with $\rho_T = m_eN_{e0} + m_pN_{p0}$ is the
net electron-positron mass density. Equation (\ref{15}) represents the general linear dispersion relation for the obliquely propagating high-frequency
EP magnetoacoustic (EPMA) mode. It is clear from equation (\ref{15}) with $``+"$ sign that for parallel
propagation $(\theta = 0^{\circ})$, the fast EPMA mode propagates as EP-Alfv\'{e}n
mode: $\omega^2/k^2 = V_A^2 = P_B /\rho_T$ in which the magnetic pressure $P_B$
gives rise to the restoring force, and the net EP mass density
$\rho_T$ provides the inertia. In stark contrast, however, when $\theta=90^{\circ}$ (i.e., perpendicular
propagation), the fast EPMA mode propagates as EP-magnetosonic mode: $\omega^2/k^2 = V_A^2(1 + \beta) = (P_B + P_T )/\rho_T$ in which
the sum of the magnetic pressure and net EP-thermal pressure
gives rise to the restoring force and the net EP mass density
provides inertia.
\begin{figure}[!t]
\centering
\includegraphics[width=6cm]{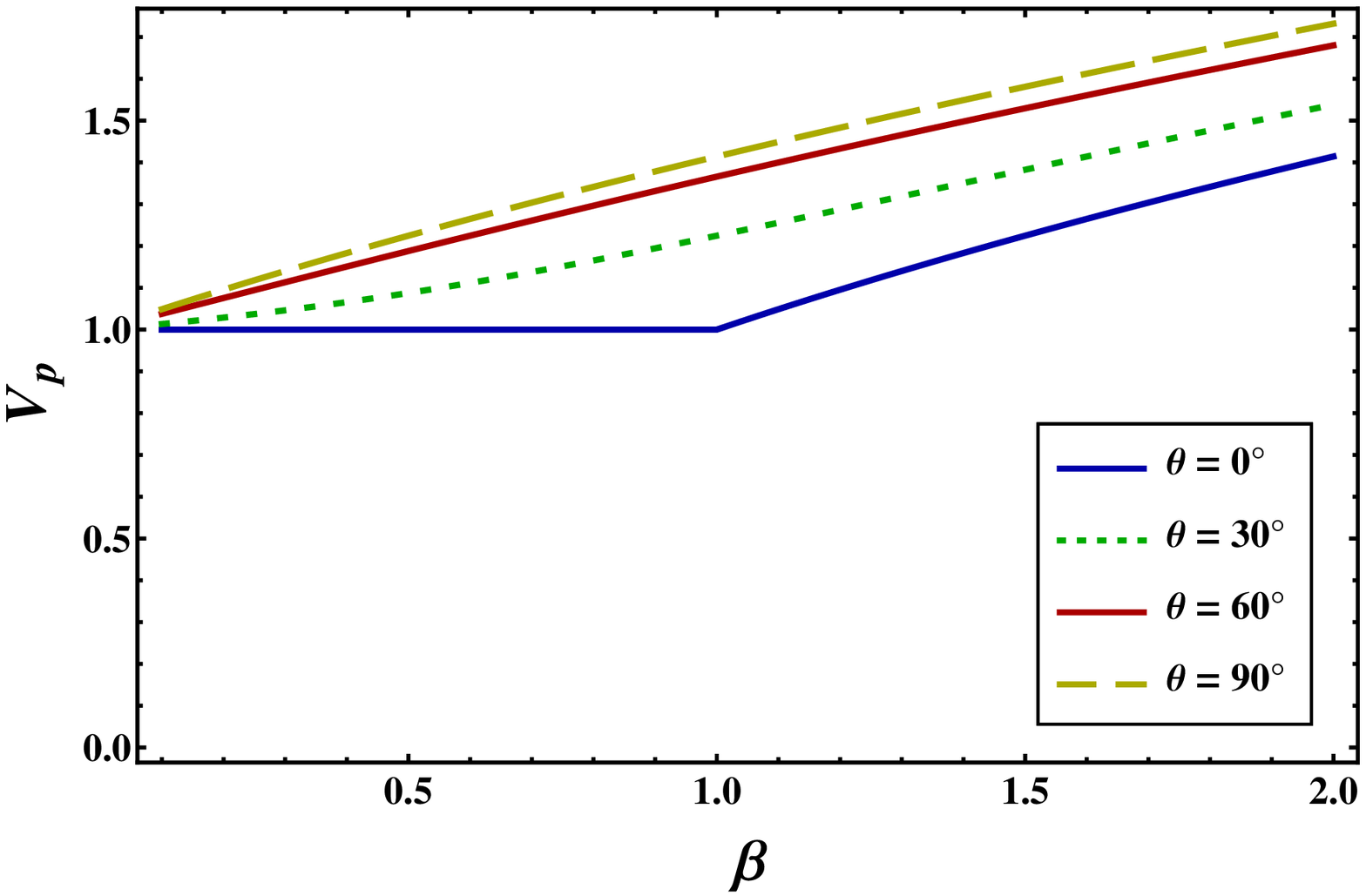}

\large{(a)}
\vspace{0.1cm}

\includegraphics[width=6cm]{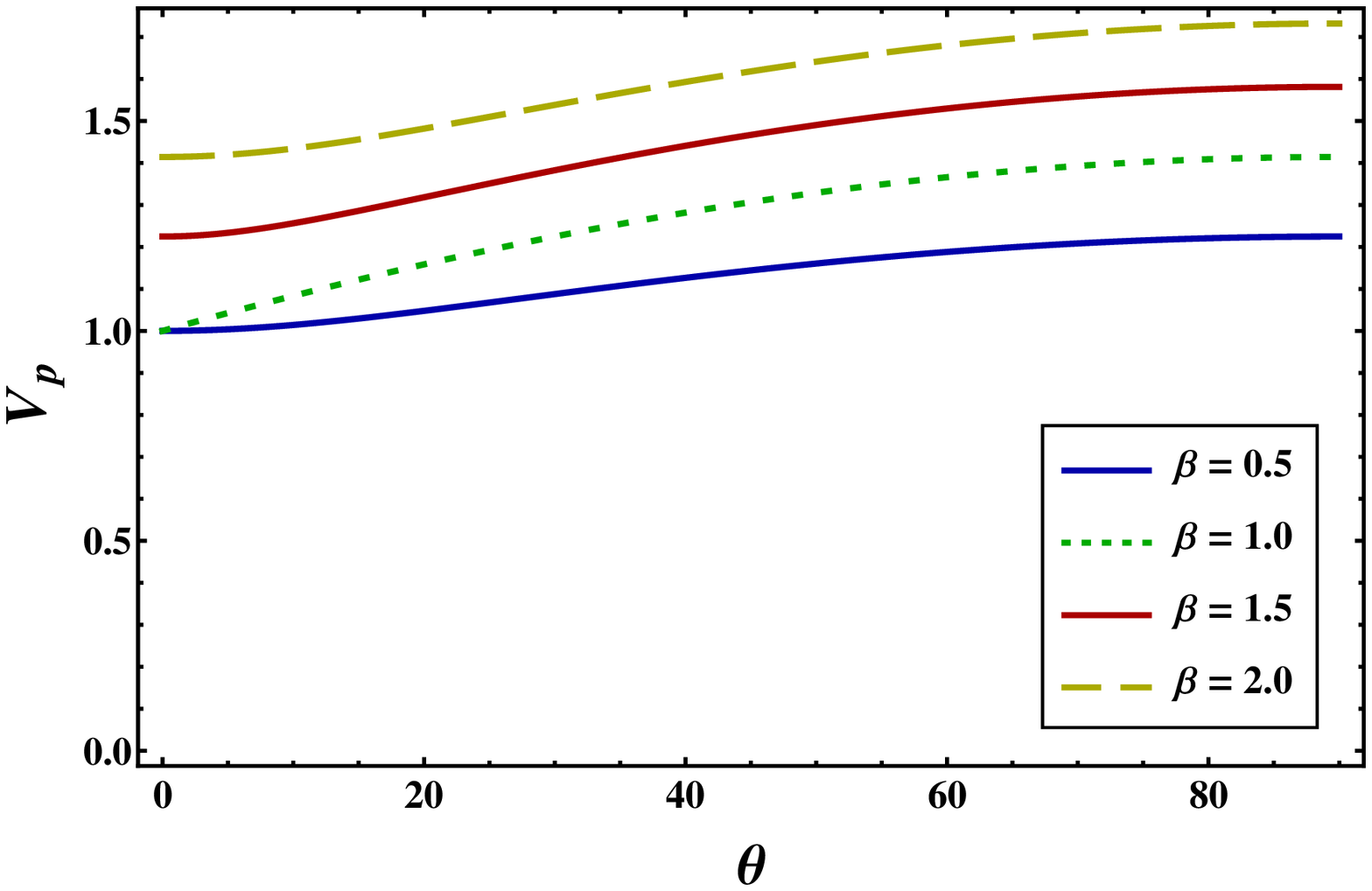}

\large{ (b)}
\caption{Showing the variation of the phase speed $V_p$ versus (a) $\beta$ for different value of $\theta$, and (b) $\theta$ for different value of $\beta$ for fast EPMA mode.} \label{Fig1}
\end{figure}

\begin{figure}[!h]
\centering
\includegraphics[width=6cm]{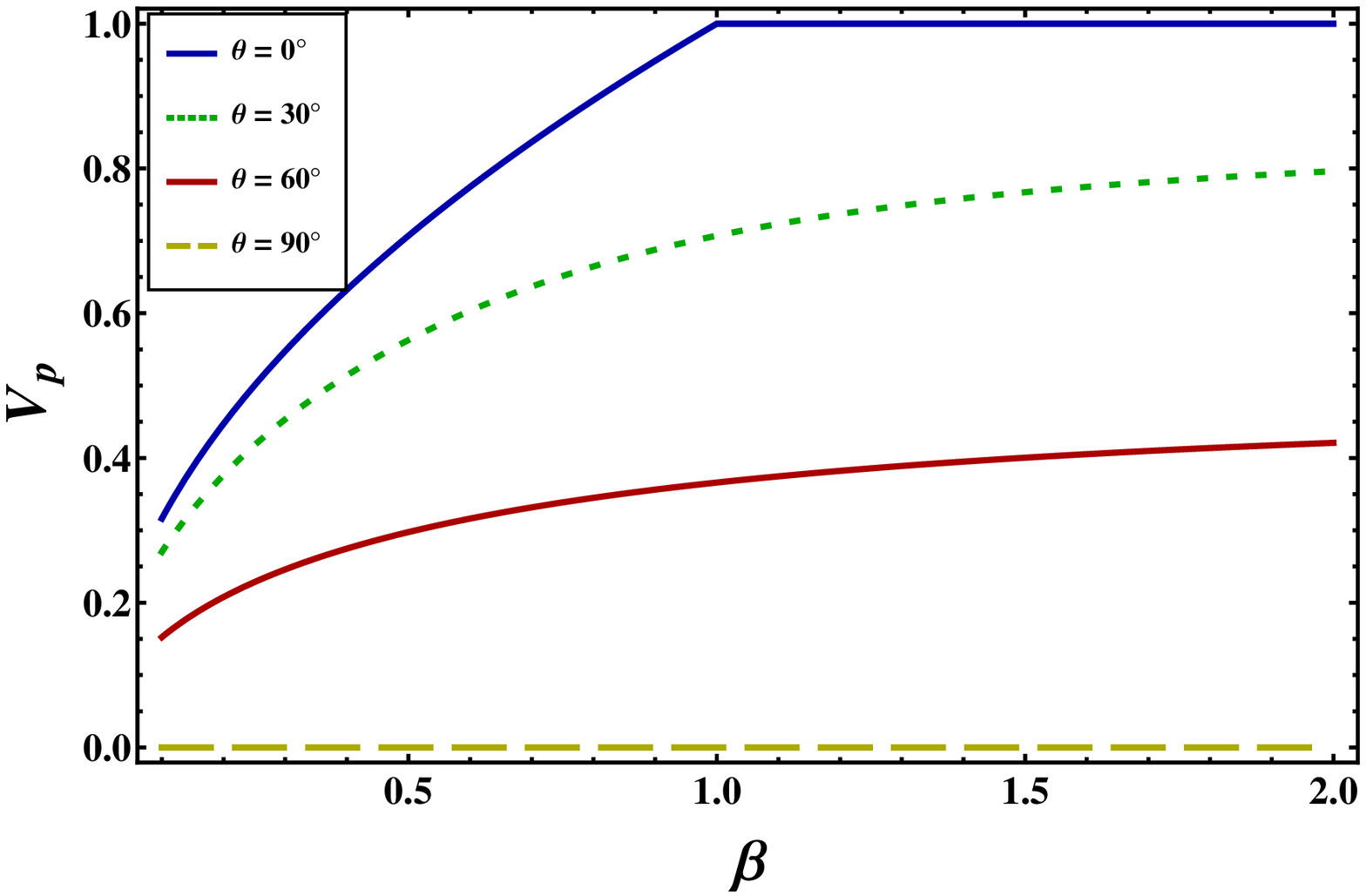}

\large{ (a)}
\vspace{0.1cm}

\includegraphics[width=6cm]{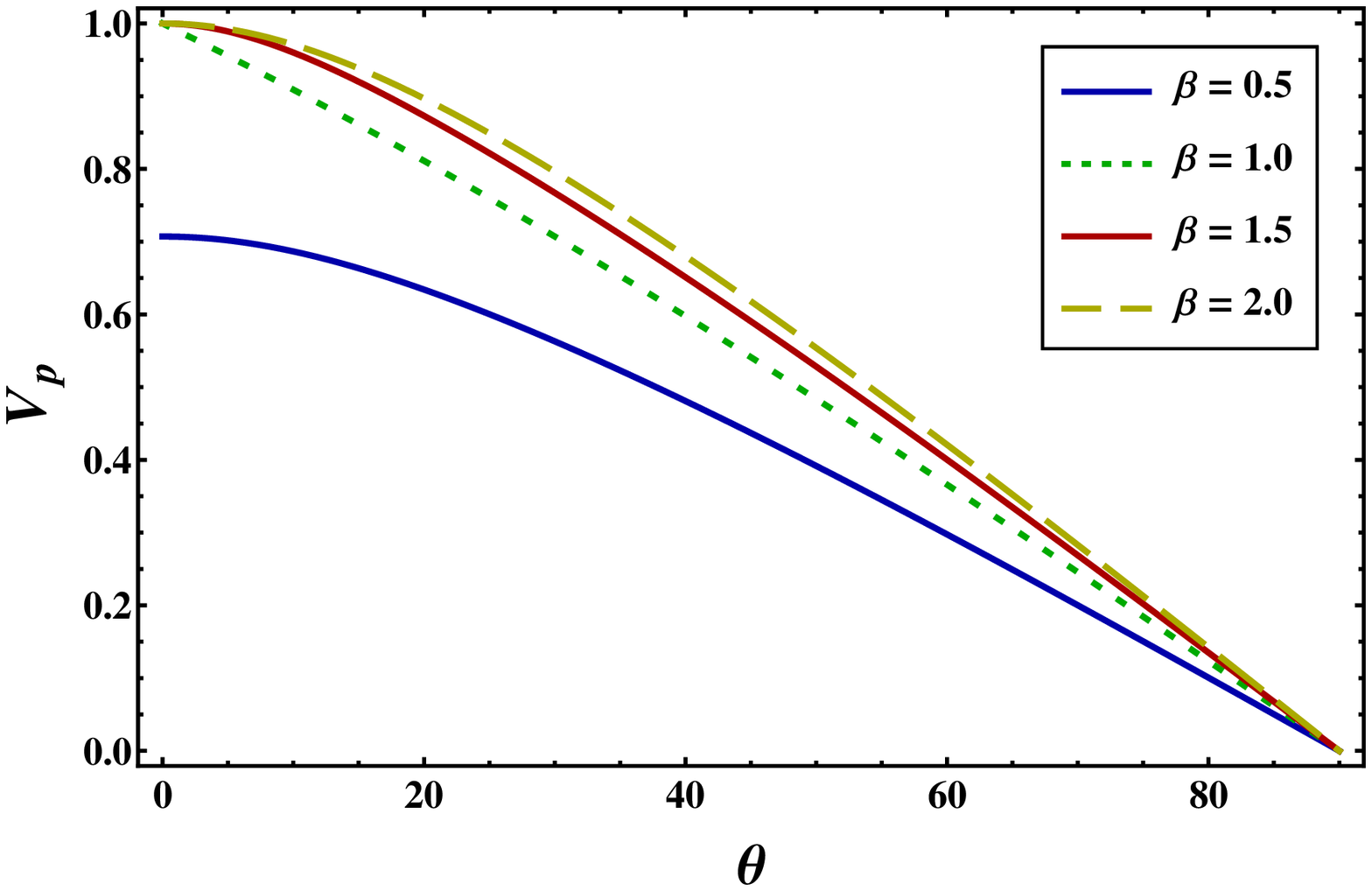}

\large{ (b)}
\caption{Showing the variation of the phase speed $V_p$ versus (a) $\beta$ for different value of $\theta$, and (b) $\theta$ for different value of $\beta$ for slow EPMA mode.} \label{Fig2}
\end{figure}

Now, substituting equations (\ref{10}) and (\ref{11}) into equations
(\ref{7})-(\ref{9}), and equating the coefficients of
$\epsilon^{2}$,  we obtain
\begin{eqnarray}
&&\hspace*{7mm}\hspace*{15mm}u_{sx}^{(1)}=1,\label{16}\\
&&\hspace*{7mm}\hspace*{15mm}b_{x}^{(1)}=0,\label{17}
\end{eqnarray}

The substitution of equations (\ref{10}) and (\ref{11}) in equations (\ref{7})-(\ref{9}), and equating the coefficients of $\epsilon^{5/2}$, we get another set of equations, which along with equations (\ref{12})-(\ref{17}), reduce to a single K-dV equation to describe nonlinear solitary waves
\begin{equation}
 {\partial_\tau}\Psi+A\Psi{\partial_\xi}\Psi+D{\partial_{\xi}^3}\Psi=0, \label{18}\
\end{equation}
where we set $b_y^{(1)}=\Psi$ for simplicity. The nonlinear coefficient can be written as
\begin{eqnarray}
&&\hspace*{0cm}A=\frac{\sqrt{2}V_p^{3}[3(V_p^{2}-\beta)^2]sin\theta}{V_p^{2}(V_p^{4}-\beta^2)sin^2\theta+(V_p^{2}-\beta)^3(V_p^{2}+cos^2\theta)}, \label{a}
\nonumber\
\end{eqnarray}
and also  the dispersion coefficient can be written as
\begin{eqnarray}
&&\hspace*{0cm}D=\frac{V_p^{3}(V_p^{2}-\beta)^2}{\sqrt[3]{2}[V_p^{2}(V_p^{2}+\beta)sin^2\theta+(V_p^{2}-\beta)^2(V_p^{2}+cos^2\theta)]} \label{b},
\nonumber\
\end{eqnarray}
where $\partial^3_\xi=\partial^3/\partial\xi^3$.  Equation
(\ref{18}), which is nothing but the K-dV (Korteweg-de Vries)
equation, describes the nonlinear propagation of the fast [when $V_p$ is given by (\ref{15}) with ``+"
sign] and slow [when $V_p$ is given by (\ref{15}) with ``-" sign] magnetosonic modes in the magnetized EP plasma medium under
consideration.

The steady-state solution of the K-dV equation (\ref{18}) is
obtained by considering a moving frame (moving with speed $u_0$)
$\zeta=\xi-u_0\tau$, and applying the appropriate boundary
conditions, viz.  $\Psi\rightarrow 0$,
$d_{\zeta}\Psi\rightarrow 0$, ${d_{\zeta}^{2}\Psi}\rightarrow 0$ at ${\zeta
\rightarrow \pm\infty}$. After some algebraic calculations
(details can be found in Ref. \cite{Karpman1975}), one can express
the steady-state solitonic solution of this K-dV equation as
\begin{eqnarray}
&&\hspace*{10mm}\Psi=\Psi_0 {sech}^2\bigg[\frac{\zeta}{\Delta}\bigg],
\label{19}
\end{eqnarray}
where the maximum potential amplitude or height $\Psi_0=3u_0/A$ (normalized by $B_0$) and the width or thickness
$\Delta=\sqrt{4D/u_0}$ (normalized by $\delta_s$) of the solitonic profile, are given by
\begin{eqnarray}
&&\hspace*{-0.5cm}\Psi_0=\frac{3u_0[V_p^{2}(V_p^{4}-\beta^2)sin^2\theta+(V_p^{2}-\beta)^3(V_p^{2}+cos^2\theta)]}{\sqrt{2}V_p^{3}[3(V_p^{2}-\beta)^2]sin\theta},
\nonumber\\
&&\hspace*{-0.5cm}\Delta=\sqrt{\frac{\sqrt{2}V_p^{3}(V_p^{2}-\beta)^2}{u_0[V_p^{2}(V_p^{2}+\beta)sin^2\theta+(V_p^{2}-\beta)^2(V_p^{2}+cos^2\theta)]}}.
\nonumber\
\end{eqnarray}

%


\begin{figure}[!t]
\centering
\includegraphics[width=6cm]{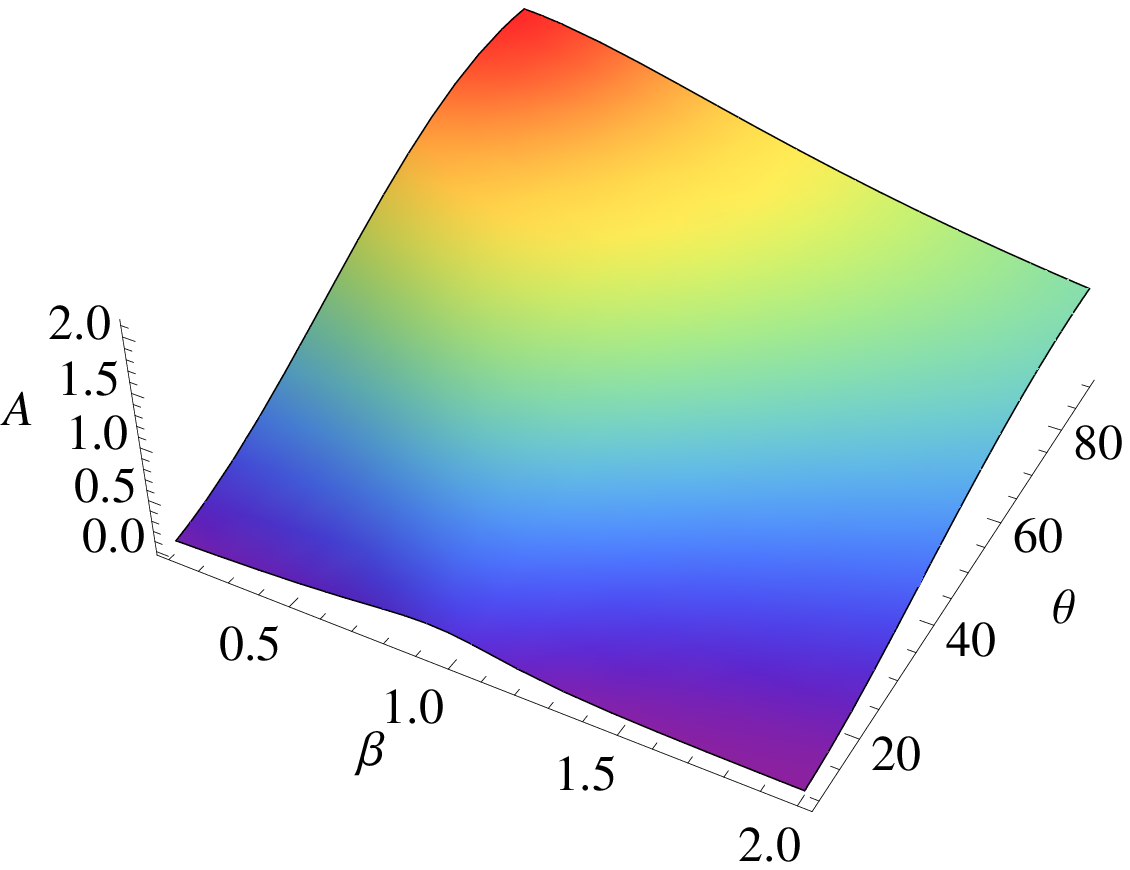}

\large{ (a)}
\vspace{0.0cm}

\includegraphics[width=6cm]{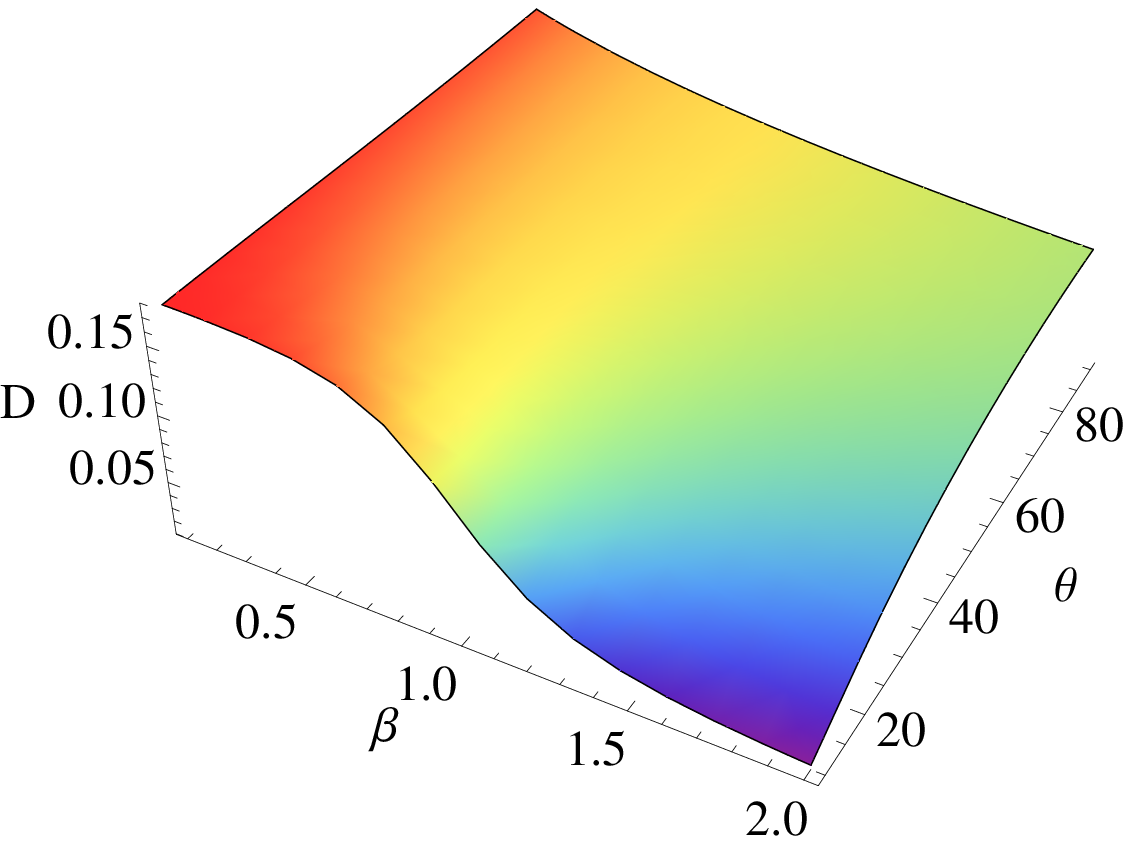}

\large{ (b)}
\caption{Showing the variation of (a) the nonlinear term $A$, and (b) the dispersion term $D$ versus $\beta$ and $\theta$ for fast (compressive) EPMA waves.} \label{Fig3}
\end{figure}

\begin{figure}[!t]
\centering
\includegraphics[width=6cm]{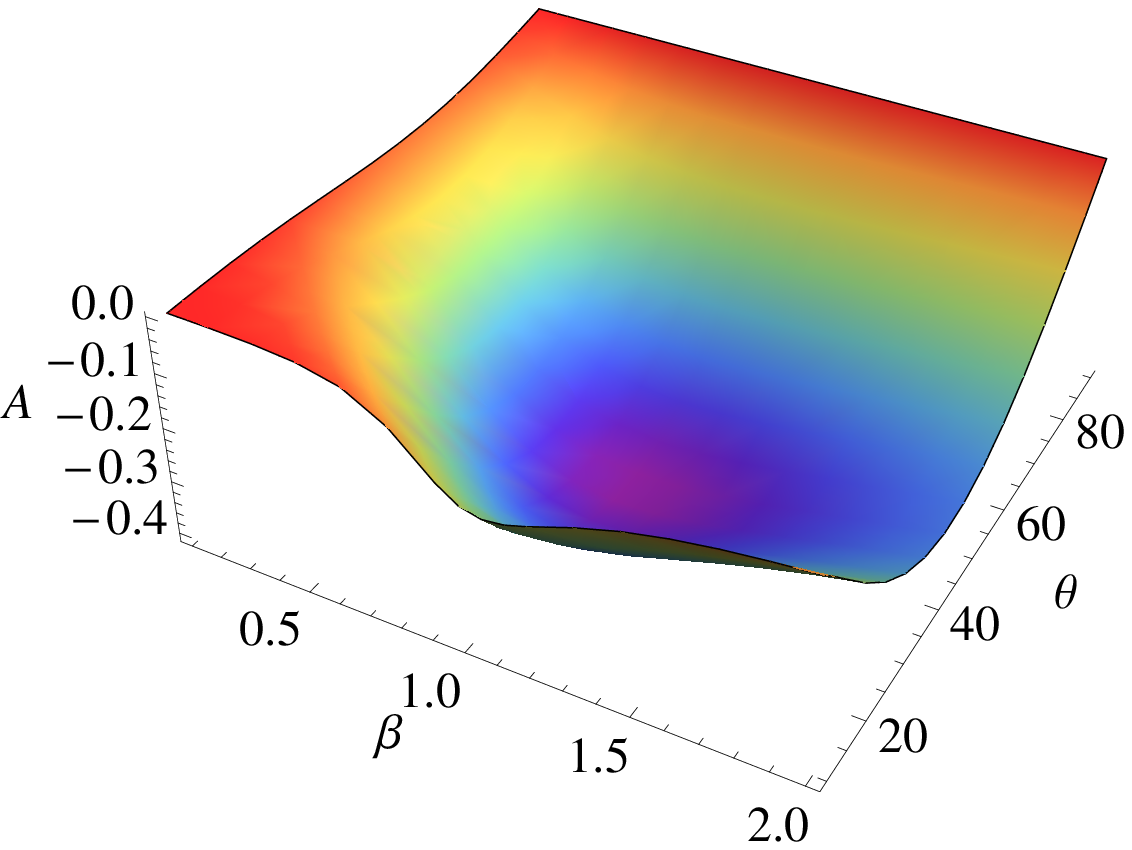}

\large{ (a)}
\vspace{0.0cm}

\includegraphics[width=6cm]{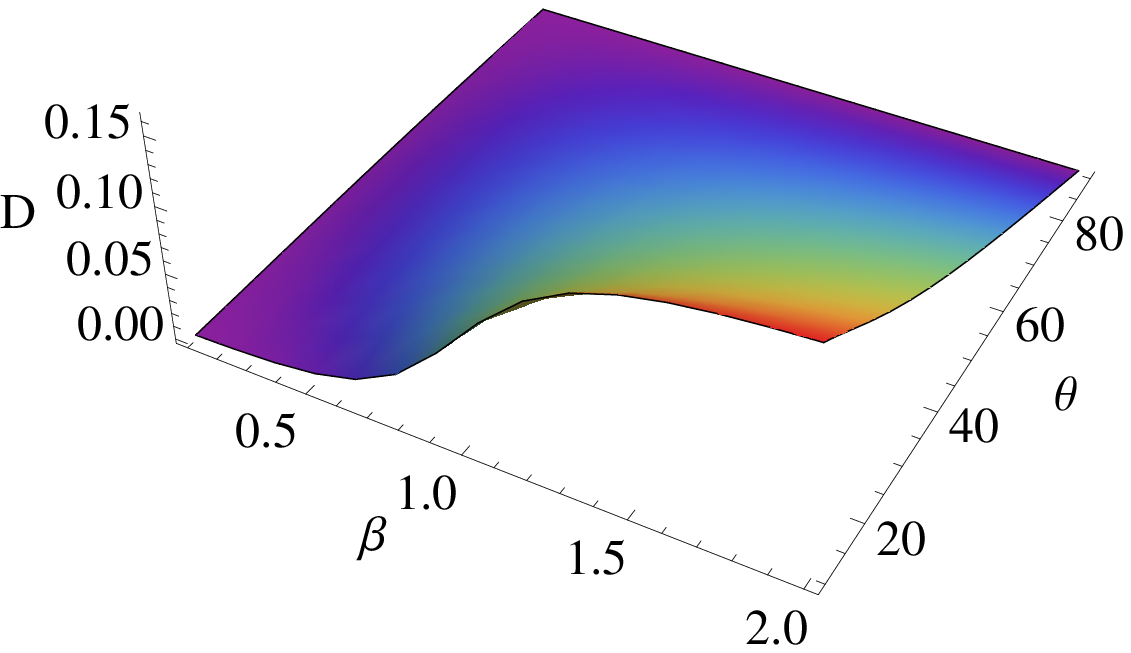}

\large{ (b)}
\caption{Showing the variation of (a) the nonlinear term $A$, and (b) the dispersion term $D$ versus $\beta$ and $\theta$ for slow (rarefactive) EPMA waves.} \label{Fig4}
\end{figure}

\begin{figure}[!t]
\centering
\includegraphics[width=6cm]{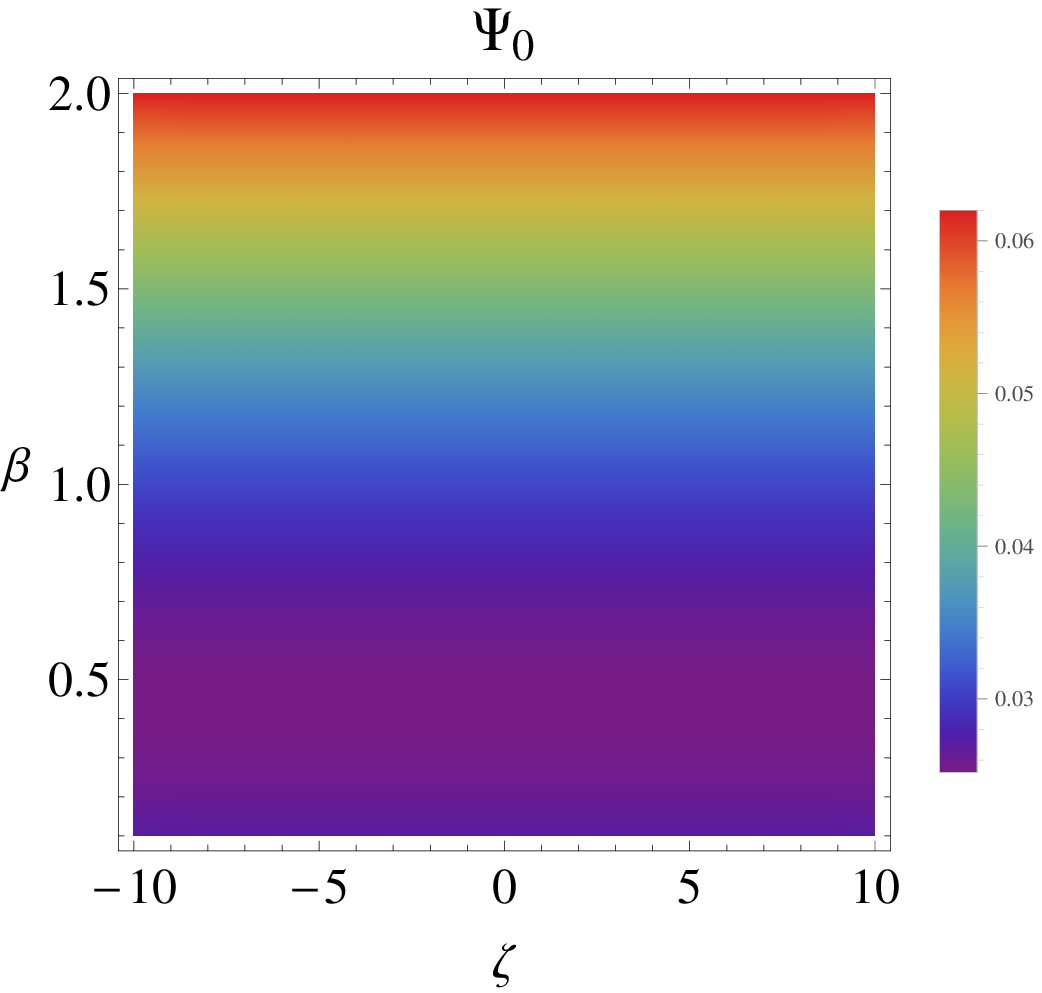}

\large{ (a)}
\vspace{0.1cm}

\includegraphics[width=6cm]{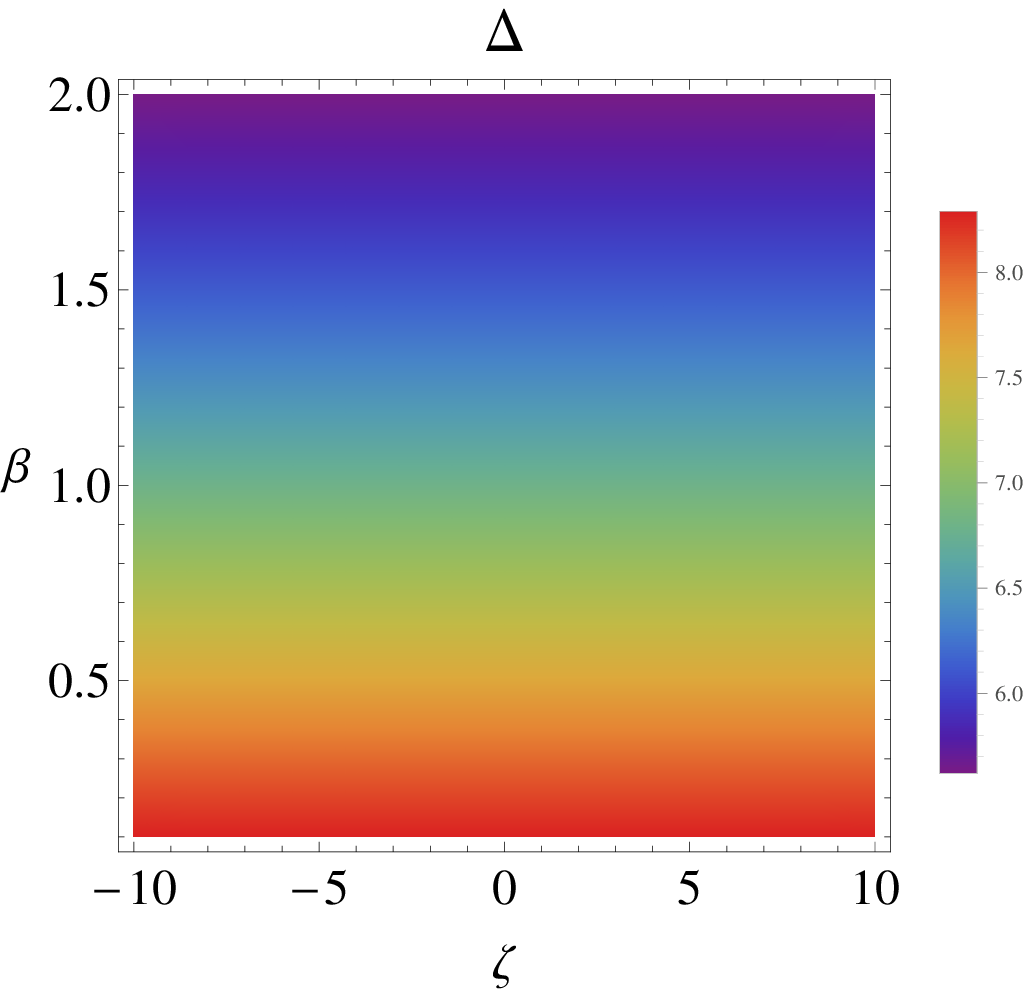}

\large{ (b)}
\vspace{0.1cm}

\includegraphics[width=6cm]{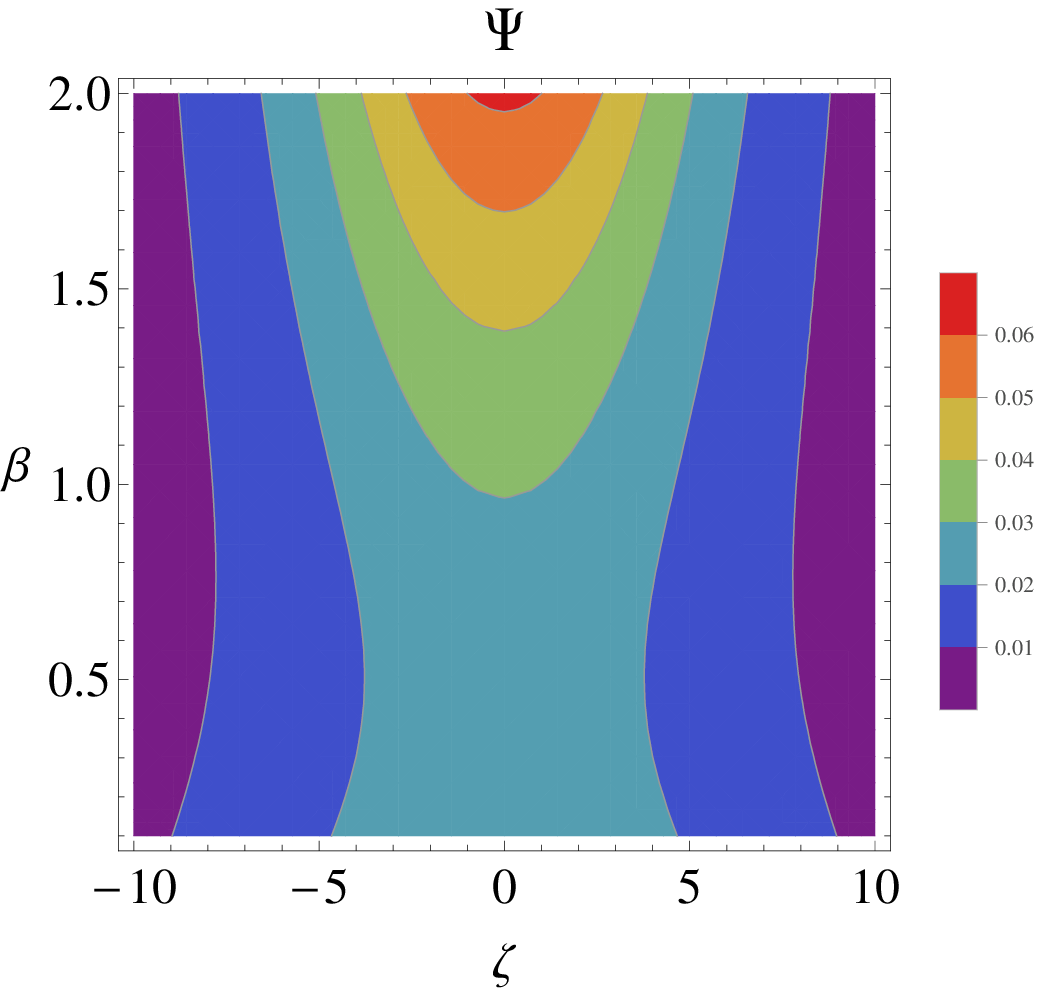}

\large{ (c)}
\caption{Showing the maximum height (a), width (b), and solitary profile (c) of the fast (compressive) EPMA waves
 along with the variation of $\zeta$ and $\beta$
with $u_0=0.01$ and $\theta=45^{\circ}$.} \label{Fig5}
\end{figure}

\begin{figure}[!t]
\centering
\includegraphics[width=6cm]{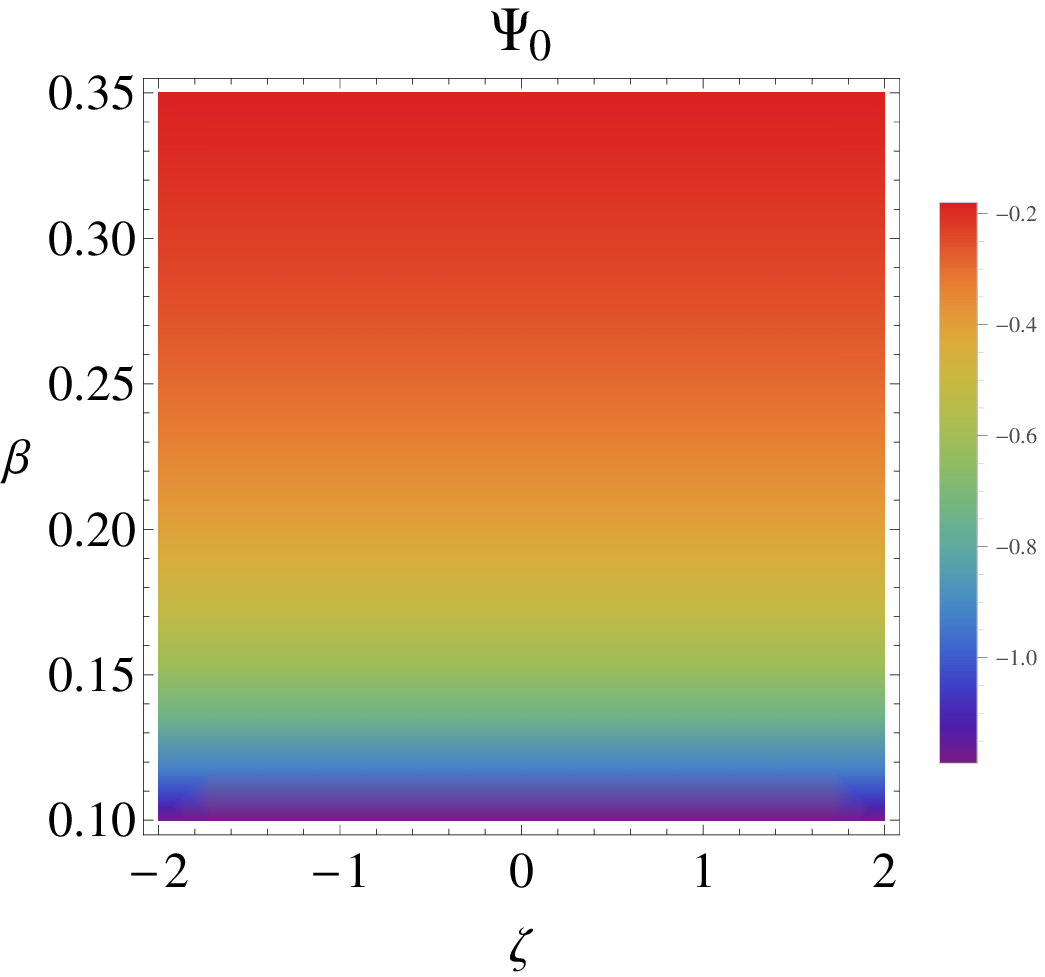}

\large{ (a)}
\vspace{0.1cm}

\includegraphics[width=6cm]{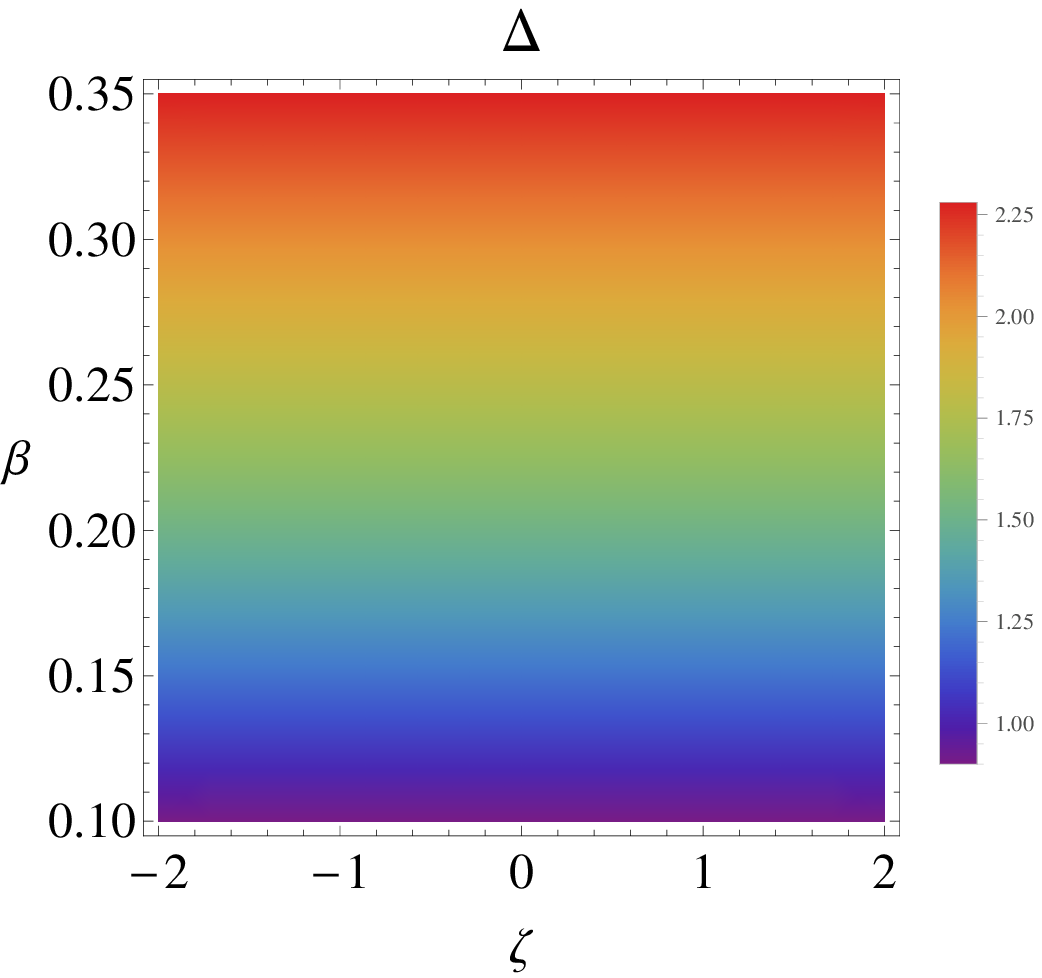}

\large{ (b)}
\vspace{0.1cm}

\includegraphics[width=6cm]{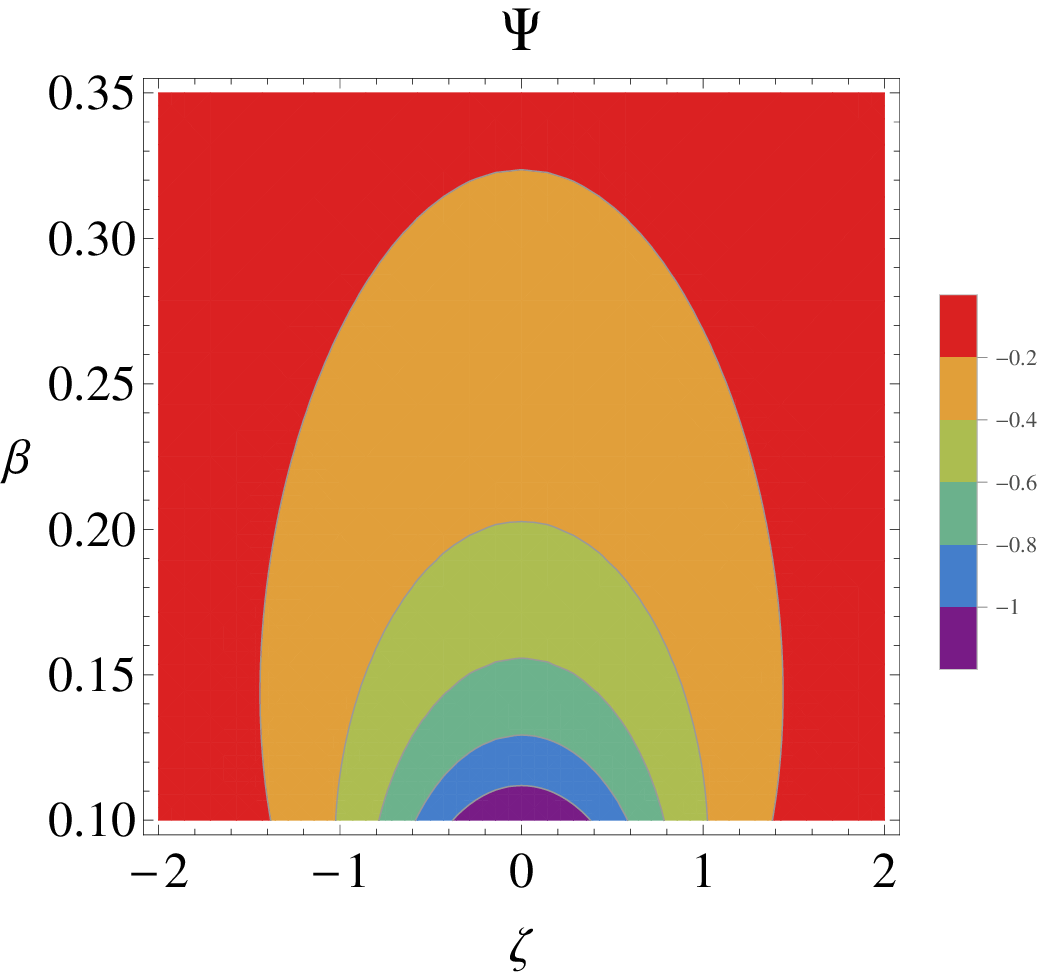}

\large{ (c)}
\caption{Showing the maximum height (a), width (b), and solitary profile (c) of the slow (rarefactive) EPMA waves
 along with the variation of $\zeta$ and $\beta$
with $u_0=0.01$ and $\theta=45^{\circ}$.} \label{Fig6}
\end{figure}

\section{Numerical Analysis and Results}
\label{NA}
We consider the propagation of electromagnetic perturbations in a pair EP plasma medium of opposite polarity of the same mass. We have numerically analyzed the phase speed $V_p$, nonlinear coefficient $A$, dispersive coefficient $D$, solitary profile in both fast and slow mode of EPMA waves.

We have numerically examined the dispersion relation for
the obliquely propagating fast EPMA mode of equation (\ref{15}) with $``+"$ sign for $0\leq\theta\leq90^{\circ}$ and $0.1\leq\beta\leq2$ in Figure \ref{Fig1}, which, in fact, graphically represents how the phase speed $V_p$ of the fast EPMA mode is modified by the effects of $\beta$ and $\theta$. It is seen that as we increase $\theta$, $V_p$ of the fast EPMA mode increases and higher the value of $\beta$ higher the value of $V_p$ with the similar value of $\theta$. However, as we increase $\beta$, $V_p$ remains constant until $\beta=1$ i.e., $V_p=1$ for $0.1\leq\beta\leq1$ at $\theta=0^{\circ}$, but it
rises with $\beta$ for $\beta>1$. In addition, it is observed that for the
fast EPMA mode $V_p>\beta$ is always satisfied.

It is obvious from equation (\ref{15}) with $``-"$ sign that the slow EPMA mode
vanishes, i.e., $V_p = 0$ for perpendicular propagation $(\theta = 90^{\circ})$ (see Figure \ref{Fig2}),
but it exists for $0\leq\theta<90^{\circ}$. The phase speed $V_p$ of the slow EPMA
mode increases as $\theta$ decrease with respect to $\beta$, and it peaked i.e., $V_p=\beta$ or $\sqrt{2}V_0=\beta$
when $\theta=0^{\circ}$. This means that this maximum phase speed increases with $\beta$. We have also numerically examined the dispersion relation for
the obliquely propagating slow EPMA mode of equation (\ref{15}) with $``-"$ sign for $0\leq\theta<90^{\circ}$ and $0.1\leq\beta\leq2$ in Figure \ref{Fig2}, which, in fact, graphically represents how the phase speed $V_p$ of the slow EPMA mode is modified by the effects of $\beta$ and $\theta$. It is found that as we increase $\theta$, $V_p$ of the slow EPMA mode decreases, however, as we increase $\beta$, $V_p$ increases for $\beta<1$, but it remains invariant for $\beta\geq1$ i.e., $V_p=1$ for $\beta\geq1$. Moreover, it is observed that for the slow EPMA mode $V_p\leq\beta$ is always satisfied.

It is obvious from that (i). EPMA solitary structures exist
if and only the dispersion coefficient is positive, i.e., $D>0$,
and (ii). EPMA solitary structures are compressive (rarefactive) if the
nonlinear coefficient $A$ is positive (negative), i.e., $A>0$ $(A<0)$
or $V_p>\beta$ $(V_p<\beta)$. We have already shown that $V_p>\beta$ $(V_p<\beta)$
is always valid for the fast (slow) EPMA mode in Figure \ref{Fig1} (Figure \ref{Fig2}). This means that
the fast (slow) EPMA mode nonlinearly propagates as compressive
(rarefactive) solitary waves.

The variation of nonlinear coefficient $A$ and dispersion coefficient $D$ is depicted in Figures \ref{Fig3} (a,b) and \ref{Fig4} (a,b) for fast (compressive) and slow (rarefactive), respectively. Now, (i) for the compressive case \ref{Fig3} (a,b), it is seen that $A$ is increased fastly (slowly) for the lower (higher) value of $\beta$, and it peaked when $\beta\leq0.5$ and $75^{\circ}\leq\theta<90^{\circ}$. On the other hand, dispersion coefficient $D$ remains invariant until $\beta=0.5$, and after that, it declines rapidly until $\beta=1$ as well as it also remains constant from $1\leq\beta\leq2$ for $\theta<20^{\circ}$. Moreover, both $A$ and $D$ go up with the increase of obliqueness $\theta$. (ii) for the rarefactive case \ref{Fig4} (a,b), nonlinear coefficient $A$ is negative here. It decreases (increases) gradually until $\beta=1$ ($\beta=2$). The same variation is also seen for $\theta$. It decreases (increases) gradually until $\theta=45^{\circ}$ ($\theta=90^{\circ}$). Interestingly, $A$ gets the maximum value when $\beta=1$ and $\theta=45^{\circ}$. In addition, the opposite scenario has been seen for dispersion coefficient $D$ compared with the compressive case.

The solitary profile is formed due to the balance between nonlinear coefficient $A$ and dispersion coefficient $D$. We have also graphically shown how the other basic features (viz. amplitude and width) of these compressive (Figure \ref{Fig5}) and rarefactive (Figure \ref{Fig6}) are modified by the effect of $\beta$ at $\theta=45^{\circ}$ with space variable $\zeta$. For example, it is observed from Figure \ref{Fig5} that the amplitude (width) of the fast (compressive) EPMA solitary structures increase (decrease) with $\beta$. On the opposite hand, Figure \ref{Fig6} indicates that the height of the slow (rarefactive) EPMA solitary structures increases at $\beta=0.1$; after that it remains invariant. Moreover, width rises due to the increase of $\beta$. In addition, a sharp solitary profile is formed that is limited by $\beta\leq0.32$ at $\zeta=|1.4|$.



\section{Discussion}
\label{D}
We have considered an EP system collisionless magnetized EP plasma to study the nonlinear propagation of the fast and slow
waves for which the mass ratio of positron-to-electron $\alpha$ is equal to one. The following are the investigation's predicted findings:

\begin{itemize}
\item{The EP medium under consideration
supports extremely high phase speed, high-frequency slow EPMA, and
fast EPMA waves (propagating, respectively, parallel and perpendicular
to the external magnetic field), in which the magnetic pressure
gives rise to the restoring force, and the net EP mass density
provides the inertia.}

\item{The phase speed of the fast (slow) EPMA mode increases (decreases)
with the increase of the obliqueness $\theta$ but increases
with the increase of the EP-thermal pressure for $\beta>1$ $(\beta <1)$ depicted in Figures \ref{Fig1} and \ref{Fig2}}.

\item{The EP plasma medium under consideration supports a
compressive (rarefactive) EPMA solitary profile that is associated
with obliquely propagating fast (slow) EPMA mode.}

\item{Hump and dip shape solitary profiles are observed for compressive and rarefactive waves, respectively. For the fast EPMA case, both the nonlinear term $A$ and the dispersive term are positive, i.e., $A,D>0$. However, for slow EPMA case, $A$ is negative but $D$ is positive i.e., $A<0$ and $D>0$. Both $A$ and $D$ increase with $\theta$, but $A$ is almost constant, and $D$ declines with $\beta$ in the case of fast EPMA mode (Figure \ref{3}). On the other hand, $A$ has the maximum value at $\beta=1$ and $\theta=45^{\circ}$ and $D$ go up (go down) with the increase of $\beta$ $(\theta)$ (Figure \ref{4}).           }

\item{The height (width) of the fast EPMA hump shape solitary waves
increases (decreases) with the increase of $\beta$ with a fixed value of $\theta$. In stark contrast, however, in the
case of slow EPMA dip shape solitary profile, as we increase $\beta$, their height (width) decreases (increases) with a fixed value of $\theta$}.
\end{itemize}

We note that the nonlinear analysis presented here is not valid when $\theta=0^{\circ}$ (exact parallel propagation), in which case one should
derive the derivative nonlinear Schr\"{o}dinger equation, and examine
the properties of the shear Alfv\'{e}n solitons \cite{Verheest1995,Mamun1999,Rajib2017}. But, except for the exact
parallel propagation $\theta=0^{\circ})$ \cite{Rajib2018}, our theory is valid for arbitrary
values of $\beta$ and $\theta$. This means that our present work is not only valid for
the opposite polarity EP plasma medium, but also valid for any kind of
two-component plasma systems, particularly dust
($\alpha=Z_nm_p/Z_pm_n$) and electron-ion ($\alpha\le 1/1836$) plasmas.

By way of conclusion, we may say that our theoretical investigation should be helpful in a better understanding of the characteristics of small but finite amplitude  electromagnetic disturbances that are ubiquitous in a laboratory as well as space plasmas, where opposite polarity plasma components are available such as the solar wind \cite{Clem2000,Adriani2009,Adriani2011}, the magnetosphere of Earth \cite{Ackermann2012,Aguilar2014}, pulsars magnetosphere \cite{Profumo2012}, and microquasars \cite{Siegert2016}.

\section*{ACKNOWLEDGMENTS}
This research article is dedicated to the little princess, \textbf{\textit{Reeha Tanvir's first birthday is on December 26, 2021}}, to keep the day memorable. One of the authors, T. I. Rajib is grateful to Jahangirnagar University (JU) for the financial support through the JU Research Grant (2021-2022). This fund has given him the freedom to focus on his research. Moreover, we acknowledge Prof A A Mamun for his continuous support.

\section*{DATA AVAILABILITY}
Data sharing is not applicable to this article as no new data were created or analyzed in this study.


\end{document}